\documentclass{article}
\usepackage{amssymb}
\usepackage{amsfonts}
\usepackage{amsmath}
\usepackage[doublespacing]{setspace}

\input{tcilatex}

\begin{document}

\title{On Dirac equation for a Coulomb scalar, vector, and tensor interaction%
}
\author{Omar Mustafa \\
Department of Physics, Eastern Mediterranean University, \\
G Magusa, North Cyprus, Mersin 10,Turkey\\
E-mail: omar.mustafa@emu.edu.tr, \\
\ Tel: +90 392 630 1314,\\
\ Fax: +90 392 3651604}
\maketitle

\begin{abstract}
In their recent paper (Inter. J. Mod. Phys. A \textbf{26} (2011) 1011),
Zarrinkamar and coauthors have considered the radial Dirac equation for a
Coulomb scalar, vector and tensor interaction. The exact solutions for the
energy eigenvalues they have reported for spin-symmetry and pseudo-spin
symmetry were mishandled and are incorrect eigenvalues, therefore.


\end{abstract}

.In their recent paper, Zarrinkamar, Hassanabadi and Rajabi (Inter. J. Mod.
Phys. A 26 (2011) 1011), have considered the radial Dirac equation for a
Coulomb scalar, vector and tensor interaction. They were successful in
working out the exact solutions as inferred from the 1D-radial Schr\"{o}%
dinger Coulomb problem as%
\begin{equation}
E^{2}-m^{2}=-\frac{b^{2}\left( E+m\right) ^{2}}{4N^{2}}\text{ ; }%
N=n_{r}+k-a+1,
\end{equation}%
for the spin symmetric case (see their Eq.(12a)), and%
\begin{equation}
E^{2}-m^{2}=-\frac{b^{2}\left( E-m\right) ^{2}}{4\,\tilde{N}^{2}}\text{ ; }%
\tilde{N}=n_{r}+k-a,
\end{equation}%
for the pseudo-spin symmetric case (see their Eq.(17a)).

A proiri, these two equations are not quadratic in $E$ (cf. e.g., Mustafa
[1,2]) to imply two branches of eigenvalues. obviously, equation (1) yields%
\begin{equation}
E-m=-\frac{b^{2}\left( E+m\right) }{4N^{2}}\text{,}
\end{equation}%
and equation (2) implies%
\begin{equation}
E+m=-\frac{b^{2}\left( E-m\right) }{4\,\tilde{N}^{2}}\text{.}
\end{equation}%
Therefore, the energy eigenvalues should read%
\begin{equation}
E=\frac{\left( 4N^{2}-b^{2}\right) m}{4N^{2}+b^{2}},
\end{equation}%
for the spin symmetric case, and%
\begin{equation}
E=-\frac{\left( 4\,\tilde{N}^{2}-b^{2}\right) m}{4\,\tilde{N}^{2}+b^{2}},
\end{equation}%
for the pseudo-spin case. 

As such, their results reported in their equations (12b) and (17b) are
unfortunate and incorrect therefore.

\end{document}